\documentclass[aps,prd,preprint,onecolumn,titlepage,eqsecnum]{revtex4-2}
\usepackage{amsmath}
\usepackage{color}
\usepackage{graphicx}
\usepackage{hyperref}
\usepackage{multirow}

\newcommand{\bq}{\begin{eqnarray}}
\newcommand{\eq}{\end{eqnarray}}
\newcommand{\bqn}{\begin{eqnarray*}}
\newcommand{\eqn}{\end{eqnarray*}}
\newcommand{\bqs}{\begin{subequations}}
\newcommand{\eqs}{\end{subequations}}
\newcommand{\bw}{\begin{widetext}}
\newcommand{\ew}{\end{widetext}}

\begin{document}
%%%%%%%%%%%%%%%%%%%%%%%%%%%%%%%%%%%%%%%%%%%%%%%%%%%%%%%%%%%%%%%%%%%%%%%%%%%%%%
%%%%%%%%%%%%%%%%%%%%%%%%%%%%%%%%%%%%%%%%%%%%%%%%%%%%%%%%%%%%%%%%%%%%%%%%%%%%%%
%%%%%%%%%%%%%%%%%%%%%%%%%%%%%%%%%%%%%%%%%%%%%%%%%%%%%%%%%%%%%%%%%%%%%%%%%%%%%%
\title{A modest redirection of quantum field theory\\ solves all current problems}

\author{John R. Klauder}
\email{klauder@ufl.edu}
\affiliation{Department of Physics and Department of Mathematics, 
University of Florida, Gainesville, USA}
\author{Riccardo Fantoni}
\email{riccardo.fantoni@posta.istruzione.it}
\affiliation{Department of Physics, University of Trieste, 
strada Costiera 11, 34151 Grignano (Trieste), Italy}

\date{\today}

\begin{abstract}
Standard quantization using, for example, path integration of field theory models, 
includes paths of momentum and field reach infinity in the Hamiltonian density, 
while the Hamiltonian itself remains finite. That fact causes considerable 
difficulties. In this paper, we represent $\pi(x)$ by $k(x)/\phi(x)$. To insure 
proper values for $\pi(x)$ it is necessary to restrict $0<|\phi(x)|<\infty$ as well 
as $0\leq|k(x)|<\infty$. Indeed that leads to Hamiltonian densities in which 
$\phi(x)^p$, where $p$ can be even integers between $4$ and $\infty$. This leads 
to a completely satisfactory quantization of field theories using situations that 
involve scaled behavior leading to an unexpected, $\hbar^2/\hat{\phi}(x)^2$ which 
arises only in the quantum aspects. Indeed, it is fair to claim that this symbol 
change leads to valid field theory quantizations.  
\end{abstract}

%\pacs{...}

\maketitle
%%%%%%%%%%%%%%%%%%%%%%%%%%%%%%%%%%%%%%%%%%%%%%%%%%%%%%%%%%%%%%%%%%%%%%%%%%%%%%
\section{Introduction}
%%%%%%%%%%%%%%%%%%%%%%%%%%%%%%%%%%%%%%%%%%%%%%%%%%%%%%%%%%%%%%%%%%%%%%%%%%%%%%

Infinity is a difficult aspect of any issue in classical and quantum physics.
Interestingly, this also infects zero as part of values that run between plus 
and minus infinity. It is evident that infinity as a symbol leads to 
$\infty+\infty=\infty$ as well as $\infty\cdot\infty=\infty$.
It is evident here that zero follows the same procedure. Namely, $0+0=0$ as 
well as $0\cdot 0=0$. It is noteworthy that zero was banned from using for 1500
years long, ago, but if a symbol, for example $A$, stands for an object which is 
specific, namely a brick or a tree or a screwdriver but these are unequivalent to 
each other having no screwdriver as a zero is just as saying that book equals zero 
and therefore book equal screwdriver equal zero. A field of nature, such as a 
particle of creation, is as unique as anything else and therefore its zero value 
deserves better treatment. In fact, if zero represents nothing then it could be that 
$\phi(x)=0$ also serves as nothing and $\phi(x)=\pi(x)=0$ leads to the fact that 
different kind of fields that appear are poorly labeled. A field $\phi(x)$ can
represent a particle of nature, but $\pi(x)$ we choose to use for the time 
derivative of $\phi(x)$. $\pi(x)=0$ with no wrong interpretation, but $\phi(x)$
should not equal zero at all. Our current field theories would use 
$-\infty<\phi(x)<+\infty$. In this case field theory has gone its happy way 
certainly classically but it really upsets things during quantization.  
One manner to be useful in eliminating $\phi(x)=0$ all together is to introduce a
friendly new classical function. In particular, we introduce $k(x)=\phi(x)\pi(x)$. 
$\phi(x)$ we said can be zero and $k(x)$ then equals zero as well. But $\phi(x)=0$ 
when $\pi(x)=300$ does not fit the physics of $k(x)=0$. It is natural that we 
withhold any $\phi(x)$ equaling zero, they are abandoned from our business and 
therefore we have correct $\phi(x)$ and $\pi(x)=k(x)/\phi(x)$ and have arranged to 
have $\phi(x)$ withdrawn from our arithmetic if it is zero. Now why can that be 
useful.

Consider the field functions in a Hamiltonian density which are $\pi(x)$ and 
$\phi(x)$ such as $\pi(x)^2$ and $\phi(x)^2+g\,|\phi(x)|^p$ is a toy example for our 
analysis. It is a fact that when $\pi$ and $\phi$ reach infinity in such a 
fashion that the integral Hamiltonian can still remain finite. In other 
words a finite Hamiltonian function can include density $\phi(x)$ and $\pi(x)$ which 
reach infinity. Why would we want that? Because the quantization, most clearly in a 
path integration, has paths which actually reach infinity in such a slim way and 
therefore include fields that contribute to the Hamiltonian itself. {\bf Nature 
should not reach infinity!}. How can we take our fields into the density and retain 
them from ever reaching infinity. If that can happen, a very much should be more 
correct in its behavior than normally used.

We change variables in such a fashion that was having $k(x)=\phi(x)\pi(x)$ and now 
use our toy Hamiltonian density our functions become 
$k(x)^2/\phi(x)^2+\phi(x)^2+g\,|\phi(x)|^p$. It is clear that $\phi(x)=0$ is helpful 
here and helpful at infinity because if infinity of $\phi$ is allowed then $k(x)$ 
could not properly serve $\pi(x)$. Moreover we can have $|k(x)|<\infty$, but equal 
zero is fine, and since $|\phi(x)|$ is also less than infinity, it follows that 
$|\phi(x)|^p$ never reaches infinity as desired. Summarizing this paragraph it says 
that $\pi$ and $\phi$ can completely substitute instead $k/\phi$. There is complete 
confidence that although the standard variables for classical theory don't want to 
reach infinity and do not, that means they are fully equivalent to our choice of the 
second version of the Hamiltonian density. Now however, we have a set of classical 
variables that are guaranteed to not reach infinity in the entire Hamiltonian 
density. Clearly it would seem that those variables deserve to become quantum 
variables. Our challenge now is to accept these two new variables $\hat{k}(x)$ and 
$\hat{\phi}(x)$, along with $\hat{\pi}^\dagger(x)\neq\hat{\pi}(x)$ we accept that 
and introduce 
$\hat{k}(x)=[\hat{\pi}^\dagger(x)\hat{\phi}(x)+\hat{\phi}(x)\hat{\pi}(x)]/2=\hat{k}^\dagger(x)$.

%%%%%%%%%%%%%%%%%%%%%%%%%%%%%%%%%%%%%%%%%%%%%%%%%%%%%%%%%%%%%%%%%%%%%%%%%%%%%%
\section{Wrestling with infinities and quantization}
%%%%%%%%%%%%%%%%%%%%%%%%%%%%%%%%%%%%%%%%%%%%%%%%%%%%%%%%%%%%%%%%%%%%%%%%%%%%%%
\label{sec1}

\subsection{Nonrenormalizability in canonical field quantization}

In standard canonical quantization, we find the equation 
$[\hat{\phi}(x),\hat{\pi}(y)]=i\hbar\delta(x-y)$. Implicitly, that applies that 
$[\hat{\phi}(x),\hat{\pi}(x)]=i\hbar\infty$ which evidently leads to absolute value 
of $|[\hat{\pi}(x),\hat{\phi}(x)]|=\hbar\infty=\infty$! In addition, while 
$\hat{\pi}(x)^2+\hat{\phi}(x)^2$ is chosen as a simple toy model for Hamiltonian 
density. This has a domain of vectors that can be normalized. However the domain of 
those vectors will shrink if the expression has $|\hat{\phi}(x)|^{2+\epsilon}$ where 
$\epsilon>0$. However if $\epsilon$ reduces to zero the domain does not recover 
itself therefore that can introduce complications. Another way to state that kind of 
problem is to choose $\hat{\pi}(x)^2+\hat{\phi}(x)^2+g\,|\hat{\phi}(x)|^p$ with 
$p>2$. Once the $g$ is positive, it determines the new domain, but if $g$ goes to 
zero, the initial domain which is bigger than the case where $g$ is positive does 
not return. Issues like that complicate conventional canonical field quantization.
Initially, we now start a procedure that simplifies matters greatly. 

\subsection{Eliminating infinities through a change of variables and a scaling}

We have already found in classical formulation of the Hamiltonian density that 
fields could reach infinity conventionally, however in using the variables 
$\pi(x)=k(x)/\phi(x)$, can eliminate classical infinities in path integration. This 
is because our density contains $k(x)^2/\phi(x)^2=\pi(x)^2$. Therefore $\phi(x)$ 
should not be zero or infinity in magnitude to complicate $\pi$ itself while $k(x)$ 
cannot reach infinity again to not complicate $\pi$. In this section we treat the 
kinetic factor $\pi(x)=k(x)/\phi(x)$ as a term needed in the quantum Hamiltonian. We 
choose $\hat{\Pi}(x)^2=\hat{k}(x)[\hat{\phi}(x)]^{-2}\hat{k}(x)$ which has been 
discussed in several articles as a suitable quantum feature of the kinetic term in 
several papers already \cite{Fantoni23d}. 
It is possible to change the term 
$\hat{k}\hat{\phi}^{-2}\hat{k}$ into $\hat{\pi}^2+a\hbar^2\infty/\hat{\phi}^2$ which 
now introduces infinities again. It is greatly noteful that the kinetic term has now 
increased itself in giving a ``potential type term''. That implies that the original 
kinetic term $\hat{\pi}^2$ has introduced as well a field potential albeit infinity. 
To obtain the equation there could be simply to take the following result somewhat 
reduced by introducing $\hat{\pi}^2+a\hbar^2W^2/\hat{\phi}^2$, where $W$ will be sent 
to infinity fairly soon. Now we do some rescaling in which we take 
$W\hat{\pi}^2$ \& $W\hat{\phi}^2$. This modification leads to 
$W\hat{\pi}^2+a\hbar^2W^2/W\hat{\phi}^2$. It is clear now that all terms are 
proportional to $W$ and therefore the final story will be taking the quantity 
$W\hat{\pi}^2+a\hbar^2W/\hat{\phi}^2$. We will multiply the previous equation by $W^{-1}$ 
to remove $W$ from both terms. This leads to finally 
$\hat{\pi}^2+a\hbar^2/\hat{\phi}^2$ which permits us to say the result can be any 
factor $A$ for the potential namely, $\hat{\pi}^2+A/\hat{\phi}^2$, in which 
$A=a\hbar^2$ and we require that $0<A<\infty$. This potential, $A/\hat{\phi}^2$, is a 
kind of potential because it is not possible to reduce $A$ back to zero in doing this 
quantization relation. It is only when quantum theory sends $\hbar\to 0$ to recover 
classical theory and then in that case $A$ would disappear.

\subsection{Affine field quantization}

So far in our story we begin to close up by establishing the essence of what we have 
said. If a classical mood for a classical Hamiltonian listed as 
$H_1=\int\{[\pi(x)^2+(\nabla\phi)(x)^2+m^2\phi(x)^2]/2+g\,\phi(x)^p\}\,d^sx$ which
represents relativistic classical Hamiltonian has difficulties especially when 
dealing with $\phi^4$ which has seen multiple records using Monte Carlo studies that 
do not lead to acceptable answers. Indeed in the model 
$H_2=\int\{[\pi(x)^2+m^2\phi(x)^2]/2+g\,\phi(x)^p\}\,d^sx$, which represents our 
ultralocal model, it will certainly fail, and has in fact done so, and done more 
poorly than the model $H_1$. However that has been using canonical quantization. A 
new quantization procedure, called {\sl affine quantization}, has been able to 
rigorously solve $H_2$ and $H_1$ automatically then. Indeed a respectable number of 
Monte Carlo studies have shown that using affine quantization can lead to very 
acceptable results for $H_2$ and $H_1$. A complete story for both cases is well 
documented in reference \cite{Fantoni23d}. 

{\bf Standard field theory 
quantization has some problems such as $\phi^4_4$ as well certainly as $\phi^{24}_4$, 
stay as failures. Remarkably, if the simple factor, $a \hbar^2 / \hat{\phi} (x)^2$, 
where $0 < a < \infty$, as this article exhibits, yields valid results for quantum 
field theory. You might like to try it.}
Readers, who are interested in the details, can be made available in that strong 
article \cite{Fantoni23d}. That article contains quantizations of field theory along 
with gravity and offers a very complete version of their being solved. For those 
readers, we hope you find whatever you will need to know.  

%%%%%%%%%%%%%%%%%%%%%%%%%%%%%%%%%%%%%%%%%%%%%%%%%%%%%%%%%%%%%%%%%%%%%%%%%%%%%%
\section{Monte Carlo examination leads to correct results of our analysis}
%%%%%%%%%%%%%%%%%%%%%%%%%%%%%%%%%%%%%%%%%%%%%%%%%%%%%%%%%%%%%%%%%%%%%%%%%%%%%%
\label{sec2}

Monte Carlo calculations are rough examples of path integration. Several Monte Carlo 
examinations by R. Fantoni have supported the results that using canonical 
quantization fails while affine quantization, the variation of quantization in this 
small paper, has proved successful. This deserves to be well known. This can be read 
in Refs. \cite{Fantoni21c, Fantoni22d}.

%%%%%%%%%%%%%%%%%%%%%%%%%%%%%%%%%%%%%%%%%%%%%%%%%%%%%%%%%%%%%%%%%%%%%%%%%%%%%%
\section{Conclusions}
%%%%%%%%%%%%%%%%%%%%%%%%%%%%%%%%%%%%%%%%%%%%%%%%%%%%%%%%%%%%%%%%%%%%%%%%%%%%%%
\label{sec:conclusions}

In this paper, we have shown that a simple switch of classical variables to 
represent $\pi(x)=k(x)/\phi(x)$. These variables all live below infinity in 
magnitude as well as field values must be greater in magnitude than zero. This 
procedure has been called upon to eliminate $\phi(x)=0$ points of nature because 
zero can also complicate equations involved. As we have seen, removing $\phi(x)$ 
values when they are zero can remarkably help in the complete and correct 
quantization of common field theories.

%%%%%%%%%%%%%%%%%%%%%%%%%%%%%%%%%%%%%%%%%%%%%%%%%%%%%%%%%%%%%%%%%%%%%%%%%%%%%%

\begin{acknowledgments}
Dr. John Klauder is pleased to have technical assistance from Dr. 
Dustin Wheeler in the preparation of this manuscript. 
\end{acknowledgments}

\subsection*{Data availability}
The data that support the findings of this study are available from the 
corresponding author upon reasonable request.

\subsection*{Conflict of interest}
The authors have no conflicts to disclose.

%%%%%%%%%%%%%%%%%%%%%%%%%%%%%%%%%%%%%%%%%%%%%%%%%%%%%%%%%%%%%%%%%%%%%%%%%%%%%%
\bibliography{klauder}
%\bibliographystyle{prsty}

%%%%%%%%%%%%%%%%%%%%%%%%%%%%%%%%%%%%%%%%%%%%%%%%%%%%%%%%%%%%%%%%%%%%%%%%%%%%%%
%%%%%%%%%%%%%%%%%%%%%%%%%%%%%%%%%%%%%%%%%%%%%%%%%%%%%%%%%%%%%%%%%%%%%%%%%%%%%%
%%%%%%%%%%%%%%%%%%%%%%%%%%%%%%%%%%%%%%%%%%%%%%%%%%%%%%%%%%%%%%%%%%%%%%%%%%%%%%
\end{document}